\documentstyle[eqsecnum,aps]{revtex}
\begin{document}
\draft
\title{Generalized Field Theoretical Approach to General Relativity and
Conserved Quantities in Anti-de Sitter Spacetimes}
\author{N. Pinto-Neto\thanks{e-mail address: nelsonpn@lafex.cbpf.br}\\ and\\R. 
Rodrigues da Silva\\Centro Brasileiro de
Pesquisas F\'{\i}sicas\\Rua Xavier Sigaud, 150 - Urca\\22290-180,
Rio de Janeiro, RJ\\Brazil}
\date{\today}
\maketitle
\begin{abstract}
A new prescription to calculate the total energies and angular momenta
of asymptotically $(d+1)$-dimensional anti-de Sitter spacetimes is
proposed. The method is based on an extension of the field theoretical 
approach to
General Relativity to the case where there is an effective cosmological
constant.  A $(d-1)$-form $\Omega$ is exhibited which, when
integrated on asymptotic $(d-1)$-dimensional boundary surfaces, yields the
values of those conserved quantities.  The calculations are gauge
independent once asymptotic conditions are not violated .
Total energies and angular momenta of some known solutions in four and
five dimensions are calculated agreeing with standard results. 
\end{abstract}

\section{Introduction}

In the last years, a lot of investigations have been done on
asymptotically anti-de Sitter (AdS) spacetimes. This interest is widely
connected with the development of string theory. For instance, the
Maldacena conjecture \cite{mal} relates conformal field theories in a
$d$-dimensional space with supergravity or string theory on the product
of $(d+1)$-dimensional AdS space with a compact manifold, and this
result can be used to study the thermodynamics of AdS black holes
\cite{wit}. One important problem of AdS space concerns the definition
of mass and angular momentum of asymptotically AdS spacetimes. Many
papers have been written on this subject proposing different ways to
calculate such quantities \cite{des,tei,ids,kra,mye}. Classically,
they all agree in their final results, with the exception of the
generalized Komar mass calculated in Ref. \cite{ids}. In this paper we
will propose a new method to calculate such quantities based on the
field theoretical approach to General Relativity (GR)
\cite{des2,gri,gri2}, which is applicable to either asymptotically flat
or asymptotically AdS spacetimes for all $d \geq 2$, and discuss its
advantages and drawbacks with respect to the other methods. Our
development also clarifies some issues concerning the background metric
defined in Refs.  \cite{des2,gri,gri2}.

In the field theoretical approach to GR, the gravitational field is
treated as a spin two field $h^{\mu\nu}$ propagating with self
interaction in a background metric $\gamma_{\mu\nu}$, which is usually
considered to be Ricci flat. The self interaction and interaction of
the spin two field with other matter fields is constructed in such a
way that the background metric is never perceived: it appears only
through the combination $\sqrt{-g}g^{\mu\nu}=
\sqrt{-\gamma}(\gamma^{\mu\nu}+h^{\mu\nu})$, and this combination
satisfies exactly the Einstein's equations. In terms of $h^{\mu\nu}$
and  $\gamma_{\mu\nu}$, the Einstein-Hilbert action and Einstein's
equation can be viewed as an action and equations of motion of a spin
two field in a Ricci flat background metric. This allows one to define
a true energy-momentum tensor of the gravitational field plus matter fields
by varying the lagrangian with respect to the background metric. 
However, the theory in
terms of $h^{\mu\nu}$ aquires a gauge freedom which is the
manifestation of the invariance of GR under the action of the manifold
mapping group (MMG), and the energy momentum tensor calculated in the
way explained above is not a gauge invariant quantity. Nevertheless,
quantities like the total energy and angular momentum do not suffer
from this ambiguity once one specify asymptotic conditions.  In fact,
some calculations of total energies have been done for asymptotically
flat spacetimes \cite{gri} agreeing with well known results. Our task
is to extend this procedure to asymptotically AdS spacetimes.  This
extension, however, is not straightforward. This is because of the
presence of the cosmological constant in AdS spacetimes. One might
think that this should not pose a problem because the formalism
developped in Refs. \cite{des2,gri,gri2} are applicable to any matter
field, and a cosmological constant can be viewed, for instance, as
resulting from the energy density of some scalar field in its ground
state.  However, when we calculate the total energy of asymptotically
AdS spacetimes considering that the background metric is still Ricci
flat, we obtain preposterous results, even when we remove the energy
of the pure AdS spacetime. The resulting energy does not yield the
usual asymptotically flat results when we put the cosmological
constant to zero. This result enforces us to take as the background
metric some Einstein space satisfying ${\stackrel{\circ}{R}}_{\mu\nu}  = 2\gamma_{\mu\nu}{\stackrel{\circ}{\Lambda}} /(d-1)$. 

In general, the total energy and angular momenta are calculated by the
integration of a closed $d$-form $J$, $dJ=0$, on a d-dimensional
spacelike volume. The main result of this paper is to exhibit the
general expression of a $(d-1)$-form $\Omega$ which fulfills the
condition $J=d\Omega$ for the general case of background metrics
satisfying ${\stackrel{\circ}{R}}_{\mu\nu}  =
2\gamma_{\mu\nu}{\stackrel{\circ}{\Lambda}} /(d-1)$, hence proving that
$J$ is indeed globally exact in general. This enables one to
write those conserved quantities as asymptotic $(d-1)$-dimensional
integrals, which rends their calculations 
feasible, and clarify the conditions for their gauge invariance. It turns 
out that the $(d-1)$-form $\Omega$ does not depend neither
on $\stackrel{\circ}{\Lambda}$ nor on $d$. With the $(d-1)$-form
$\Omega$ we have at our disposal an alternative and straightforward
method to calculate any conserved quantity of any asymptotically flat
or AdS spacetime in any number of dimensions greater than two.
We calculated
total energies and angular momenta for some known solutions in four and
five dimensions using our expression for $\Omega$, and they agree with
standard results. 

This paper is organized as follows: in the next section we generalize
the field theoretical approach to the case of GR with a cosmological
constant $\Lambda$ and background metric satisfying
${\stackrel{\circ}{R}}_{\mu\nu}  =
2\gamma_{\mu\nu}\stackrel{\circ}{\Lambda} /(d-1)$ (note that, for the
sake of generality, the two cosmological constants may be different).
We obtain the total energy-momentum tensor.  In section III
we write down the conserved current $J$ associated with this total 
energy-momentum tensor, and we arrive at the $(d-1)$-form
$\Omega$. We discuss the conditions for its gauge invariance, and we
calculate some conserved quantities of well known examples using this
method. We also discuss the necessity of using background metrics
satisfying ${\stackrel{\circ}{R}}_{\mu\nu}  =
2\gamma_{\mu\nu}\stackrel{\circ}{\Lambda}/(d-1)$ when the effective
geometry is asymptotically AdS with $\stackrel{\circ}{\Lambda}$ being
the same as the effective cosmological constant in the effective
geometry.  We end up in section IV with conclusions and discussions.

\section{The field theoretical approach for General Realtivity with a
cosmological constant}

Take the Einstein-Hilbert action depending on a metric $g_{\mu\nu}$ and
a symmetric connection $\Gamma^{\alpha}_{\mu\nu}$, together with a cosmological
constant $\Lambda$ term, and an action for matter fields without
couplings with the connection. In this total action, make the substitutions

\begin{equation}
\sqrt{-g}g^{\mu\nu}= \sqrt{-\gamma}(\gamma^{\mu\nu}+h^{\mu\nu}),
\label{gh}
\end{equation}
\begin{equation}
\Gamma^{\alpha}_{\mu\nu}= C^{\alpha}_{\mu\nu}+K^{\alpha}_{\mu\nu}.
\label{CK}
\end{equation}
where $\gamma_{\mu\nu}$ and $C^{\alpha}_{\mu\nu}$ are the metric and
Christoffel symbols, respectively, of a background space satisfying 
\begin{equation}
\label{1}
{\stackrel{\circ}{R}}_{\mu\nu}  = \frac{2\stackrel{\circ}{\Lambda} }{d-1}\gamma_{\mu\nu} 
\end{equation}
From now on indices will
be raised and lowered by $\gamma_{\mu\nu}$. After discarding
surface terms and terms independent on the fields $h^{\mu\nu}$    
and $K^{\alpha}_{\mu\nu}$, we obtain:

\begin{eqnarray}
S &=& \frac{1}{16 \pi G}
\int\biggl[\tilde{h}^{\mu\nu}(K^{\alpha}_{\mu\nu;\alpha}-K_{\mu;\nu})+
(\tilde{\gamma}^{\mu\nu}+\tilde{h}^{\mu\nu})(KK)_{\mu\nu} \nonumber \\
& & + \frac{2}{d-1}\stackrel{\circ}{\Lambda} \tilde{h}^{\mu\nu}\gamma_{\mu\nu}-
2\Lambda\sqrt{-\bar{g}}\biggl]d^{d+1}x
+\int{\cal {L}}_M (\Phi_A,\bar{g}_{\alpha\beta})d^{d+1}x,
\label{S}
\end{eqnarray}
where
\begin{equation}
\tilde{\gamma}^{\mu\nu}\equiv \sqrt{-\gamma}\gamma^{\mu\nu}\,\, , 
\tilde{h}^{\mu\nu}\equiv \sqrt{-\gamma}h^{\mu\nu},
\label{DI}
\end{equation}
\begin{equation}
K_{\mu}\equiv K^{\alpha}_{\alpha\mu},
\label{DII}
\end{equation}
and
\begin{equation}
(KK)_{\mu\nu}\equiv 
K^{\alpha}_{\mu\nu}K_{\alpha}-K^{\alpha}_{\mu\beta}K^{\beta}_{\nu\alpha}.
\label{DIII}
\end{equation}
The semi-colon (;) represents the covariant derivative with respect to the background connections, and $\Phi_A$ represents general matter fields. The 
terms $\bar{g}_{\alpha\beta}$, which is the inverse of $g^{\mu\nu}$, and $\sqrt{-\bar{g}}$ appearing in the
action must be understood as algebraic 
functions of finite degree polynomials on $h^{\mu\nu}$ and $\gamma_{\mu\nu}$
(see Ref. \cite{ron} for their explicit form).  

The dynamical equations for the gravitational field can be obtained by 
performing a Palatini-like variation of the action (\ref{S}) with respect to the fields $\tilde{h}^{\mu\nu}$ and $K^{\alpha}_{\mu\nu}$ independently.
We obtain the following equations: 
\begin{eqnarray}
\label{K}
& & K^{\alpha}_{\mu\nu;\alpha} - \frac{1}{2}(K_{\mu ;\nu} + K_{\nu ;\mu})+
(KK)_{\mu\nu}=\nonumber \\
& &\frac{2}{d-1}(\Lambda\bar{g}_{\mu\nu}-\stackrel{\circ}{\Lambda} \gamma_{\mu\nu})
+8\pi G \biggl(T_{\mu\nu}^M - \frac{1}{d-1}T^M\gamma_{\mu\nu}\biggr),
\end{eqnarray}
and
\begin{equation}
\tilde{h}^{\mu\nu}_{\;\;\;\; ;\alpha}
-(\tilde{\gamma}^{\mu\nu}+\tilde{h}^{\mu\nu})K_{\alpha}+
(\tilde{\gamma}^{\mu\rho}+
\tilde{h}^{\mu\rho})K^{\nu}_{\rho\alpha}+(\tilde{\gamma}^{\nu\rho}+\tilde{h}^{\nu\rho})K^{\mu}_{\rho\alpha}= 0,
\label{hK}
\end{equation}
where
\begin{equation}
T_{\mu\nu}^M = -\frac{2}{\sqrt{-\gamma}} \frac{\delta {\cal {L}}_M}
{\delta \gamma^{\mu\nu}},
\end{equation}
and $T^M$ is its trace.
Note that the matter energy-momentum tensor defined above is not the
same as the usual matter energy-momentum tensor $t_{\mu\nu}^M$ 
defined using variations of
the matter lagrangian with respect to the effective metric $g^{\mu\nu}$.
Only the combinations
\begin{equation}
T_{\mu\nu}^M - \frac{1}{d-1} T_{\alpha\beta}^M\gamma^{\alpha\beta} 
\gamma_{\mu\nu} = -2 \frac{\delta {\cal {L}}_M}
{\delta \tilde{\gamma}^{\mu\nu}} = -2 \frac{\delta {\cal {L}}_M}
{\delta \tilde{g}^{\mu\nu}} = t_{\mu\nu}^M - \frac{1}{d-1} t_{\alpha\beta}^M g^{\alpha\beta} g_{\mu\nu}
\end{equation}
are equal. Note also that the $T_{\mu\nu}^M$ defined above is not covariantly
conserved with respect to the background connection.

Eq. (\ref{hK}) is equivalent to the determination of the connection
$\Gamma^{\alpha}_{\mu\nu}$ in terms
of the Christoffel symbols of $g_{\mu\nu}$ 
while Eq. (\ref{K}) is equivalent to the Einstein's
equations when we use Eqs. (\ref{gh},\ref{CK},\ref{1}).
After some rearrangements in Eqs. (\ref{K},\ref{hK})
(see Ref. \cite{gri} for details), we obtain the following
equation:
\begin{eqnarray}
G^{L}_{\mu\nu} &=& 
-(KK)_{\mu\nu}+\frac{1}{2}\gamma_{\mu\nu}(KK)^{\alpha}_{\alpha}+
Q^{\alpha}_{\mu\nu ;\alpha} \nonumber \\ & &+
\frac{\Lambda}{d-1}(2\bar{g}_{\mu\nu}-\bar{g}_{\alpha}^{\alpha}\gamma_{\mu\nu})
+\stackrel{\circ}{\Lambda} \gamma_{\mu\nu} + 8\pi G T^{M}_{\mu\nu},
\label{G}
\end{eqnarray}
where
\begin{equation}
2G^{L}_{\mu\nu}= 
[\gamma_{\mu\nu}h^{\alpha\beta}+\gamma^{\alpha\beta}h_{\mu\nu}-\delta^{\alpha}_
{\mu}h^{\beta}_{\nu}-\delta^{\alpha}_{\nu}h^{\beta}_{\mu}]_{;\alpha ;\beta},
\label{2G}
\end{equation}
and
\begin{equation}
\begin{array}{rcl}
2Q^{\alpha}_{\mu\nu} \ &=& \ -\gamma_{\mu\nu}h^{\rho\sigma}K^{\alpha}_{\rho\sigma}+h_{\mu\nu}K^{\alpha}-h^{\alpha}_{\mu}K_{\nu}-h^{\alpha}_{\nu}K_{\mu} + \\
\hbox{} &+& \ 
h^{\rho}_{\mu}(K^{\alpha}_{\rho\nu}-
K^{\sigma}_{\rho\lambda}\gamma^{\alpha\lambda}\gamma_{\sigma\nu})+h^{\rho}_{\nu}
(K^{\alpha}_{\rho\mu}-K^{\sigma}_{\rho\lambda}\gamma^{\alpha\lambda}
\gamma_{\sigma\mu}) + \\
\hbox{} &+& \ 
h^{\alpha\rho}(K^{\sigma}_{\rho\mu}\gamma_{\sigma\nu}+
K^{\sigma}_{\rho\nu}\gamma_{\sigma\mu}),
\end{array}
\label{2Q}
\end{equation}

The total energy-momentum tensor of the gravitational field plus matter
described by the action (\ref{S}) can be calculated as usual by varying the 
total lagrangian ${\cal {L}}_T={\cal {L}}_G+{\cal {L}}_M$ with respect to the background metric, yielding
\begin{eqnarray}
T_{\mu\nu}^T &=& -\frac{2}{\sqrt{-\gamma}} \frac{\delta {\cal {L}}_T}{\delta 
\gamma^{\mu\nu}} = \frac{1}{8\pi G}\biggr[ -(KK)_{\mu\nu}+\frac{1}{2}\gamma_{\mu\nu}(KK)^{\alpha}_{\alpha}+
Q^{\alpha}_{\mu\nu ;\alpha} \nonumber \\ & & +\frac{\Lambda}{d-1}(2\bar{g}_{\mu\nu}-
\bar{g}_{\alpha}^{\alpha}\gamma_{\mu\nu})+
\frac{2\stackrel{\circ}{\Lambda}}{d-1} h_{\mu\nu}\biggl]+T_{\mu\nu}^M.
\label{t}
\end{eqnarray}
Equations (\ref{G}) and (\ref{t}) give rise to
\begin{equation}
G^{L}_{\mu\nu} + \frac{2\stackrel{\circ}{\Lambda} }{d-1} h_{\mu\nu} =
8 \pi G T_{\mu\nu}^T + \stackrel{\circ}{\Lambda}  \gamma_{\mu\nu}.
\label{tT0}
\end{equation}
Due to the Riemannian nature of the background geometry, one can define a new conseved energy-momentum tensor as
\begin{equation}
T_{\mu\nu}= T_{\mu\nu}^T + \frac{\stackrel{\circ}{\Lambda}}{8\pi G}  \gamma_{\mu\nu},
\label{tr}
\end{equation}
and write Eq. (\ref{tT0}) as
\begin{equation}
G^{L}_{\mu\nu} + \frac{2\stackrel{\circ}{\Lambda} }{d-1} h_{\mu\nu} =
8 \pi G T_{\mu\nu}.
\label{tT}
\end{equation}
There are some advantages in adopting $T_{\mu\nu}$ instead of
$T_{\mu\nu}^T$.  For instance, let us assume that the background metric
is a solution of the Einstein's equations when the matter fields are in
the vacuum state, which is equivalent to say that $h_{\mu\nu} = 0$ is a
solution of Eqs. (\ref{K}) and (\ref{hK}) in vacuum. This is a
reasonable assumption, although not mandatory at this moment. In this
case, $T_{\mu\nu}(h_{\mu\nu} = 0)$ is identically zero while
$T_{\mu\nu}^T(h_{\mu\nu} = 0)$ has a residual background term, as can be seen
imediately from Eqs. (\ref{tT0},\ref{tT}). As we
will see in the next section, the new $T_{\mu\nu}$ is a renormalized
energy-momentum tensor yielding authomatically finite results for
conserved quantities.

Note that the covariant divergence of the left-hand-side of Eq. (\ref{tT})
is identically zero once the background metric satisfies Eq. (\ref{1}).
Hence, the conservation of $T_{\mu\nu}$ can also
be obtained as a consequence of the field equations. 

As the energy-momentum tensor $T_{\mu\nu}$ is a true tensor, it seems that
it does not suffer from the ambiguities which are present in the usual
definitions of pseudo-tensors in GR. However, the field theoretical
approach to GR has an invariance under true gauge transformations that
comes from the invariance of GR under the action of the manifold mapping
group. The $T_{\mu\nu}$ above defined is not a gauge invariant quantity,
as it will now be seen.

The coordinate transformation invariance of GR is translated to invariance under gauge transformations on $h^{\mu\nu}$ and $K^{\alpha}_{\mu\nu}$ in the field
theoretical approach in the following way: consider the infinitesimal coordinate transformation $x'^{\alpha}= x^{\alpha}+\xi^{\alpha}(x)$, which changes the functional form of $\tilde{g}^{\mu\nu}$ as   
\begin{equation}
\tilde{g}'^{\mu\nu}(x)= 
\tilde{g}^{\mu\nu}(x)+\pounds^{(1)}_{\xi}\tilde{g}^{\mu\nu}(x),
\label{g'g}
\end{equation}
where the Lie derivative $\pounds^{(1)}_{\xi}\tilde{g}^{\mu\nu}$ is given by 
\begin{equation}
\pounds^{(1)}_{\xi}\tilde{g}^{\mu\nu}(x)= 
-\tilde{g}^{\mu\nu}_{,\lambda}\xi^{\lambda}+
\tilde{g}^{\lambda\mu}\xi^{\nu}_{,\lambda}+
\tilde{g}^{\lambda\nu}\xi^{\mu}_{,\lambda}-
\tilde{g}^{\mu\nu}\xi^{\sigma}_{,\sigma}.
\label{LG}
\end{equation}
In the case of a finite transformation, the change in $\tilde{g}^{\mu\nu}$ is given by
\begin{equation}
\tilde{g}'^{\mu\nu}(x)= 
\tilde{g}^{\mu\nu}(x)+\sum_{k=1}^{\infty}\frac{1}{k!}\pounds^{(k)}_{\xi}\tilde{g}^{\mu\nu},
\label{LGG'}
\end{equation}
where $\pounds^{(k)}_{\xi}$ is the Lie derivative of order $k$ defined as
\begin{equation}
\pounds^{(k)}_{\xi}= \pounds^{(1)}_{\xi}[ \pounds^{(k-1)}_{\xi}].
\label{Lk}
\end{equation}
Substituting in (\ref{LGG'}) the definition (\ref{gh}) we get
\begin{equation}
\tilde{g}'^{\mu\nu}(x)= 
\tilde{\gamma}^{\mu\nu}(x)+\tilde{h}^{\mu\nu}(x)+
\sum_{k=1}^{\infty}\frac{1}{k!}\pounds^{(k)}_{\xi}(\tilde{\gamma}^{\mu\nu}+\tilde{h}^{\mu\nu}).
\label{Lgh}
\end{equation}
The transformed metric density can be decomposed in two distinct ways:
\begin{equation}
\tilde{g}'^{\mu\nu}(x)= 
\tilde{\gamma}^{\mu\nu}(x)+\tilde{h}'^{\mu\nu}(x)
\label{gh'}
\end{equation}
and
\begin{equation}
\tilde{g}'^{\mu\nu}(x)= \tilde{\gamma}^{*\mu\nu}(x)+\tilde{h}^{*\mu\nu}(x),
\label{gh*}
\end{equation}
where, by comparison with (\ref{Lgh}) one gets 
\begin{equation}
\tilde{h}'^{\mu\nu}(x)= 
\tilde{h}^{\mu\nu}(x)+\sum_{k=1}^{\infty}\frac{1}{k!}\pounds^{(k)}_{\xi}(\tilde
{\gamma}^{\mu\nu}+\tilde{h}^{\mu\nu}),
\label{h'h}
\end{equation}
\begin{equation}
\tilde{h}^{*\mu\nu}(x)= 
\tilde{h}^{\mu\nu}(x)+\sum_{k=1}^{\infty}\frac{1}{k!}\pounds^{(k)}_{\xi}\tilde{h}^{\mu\nu},
\label{hh*}
\end{equation}
and
\begin{equation}
\tilde{\gamma}^{*\mu\nu}(x)= \tilde{\gamma}^{\mu\nu}(x)+
\sum_{k=1}^{\infty}\frac{1}{k!}\pounds^{(k)}_{\xi}\tilde{\gamma}^{\mu\nu}.
\label{gama*}
\end{equation}
Analogously, the gauge transformation corresponding to Eq. (\ref{gh'}) for the field $K^{\alpha}_{\mu\nu}(x)$ is
\begin{equation}
K'^{\alpha}_{\mu\nu}(x)= 
K^{\alpha}_{\mu\nu}(x)+
\sum^{\infty}_{k=1}\frac{1}{k!}\pounds^{(k)}_{\xi}(C^{\alpha}_{\mu\nu}+
K^{\alpha}_{\mu\nu}),
\label{LKK'}
\end{equation}
while for the case corresponding to the transformation (\ref{gh*}) reads
\begin{equation}
K^{*\alpha}_{\mu\nu}(x)= 
K^{\alpha}_{\mu\nu}(x)+
\sum^{\infty}_{k=1}\frac{1}{k!}\pounds^{(k)}_{\xi}K^{\alpha}_{\mu\nu}.
\label{LKK*}
\end{equation}

Eqs. (\ref{gh*}), (\ref{hh*}),(\ref{gama*}) and (\ref{LKK*}) represent the usual transformations on tensorial fields resulting from a general mapping of the manifold on which they are defined. Hence, all tensors in the manifold, in particular the energy-momentum tensor defined above, will transform in the 
usual homogeneous way: 
\begin{equation}
T'^{\mu\nu}(x)= 
T^{\mu\nu}(x)+
\sum^{\infty}_{k=1}\frac{1}{k!}\pounds^{(k)}_{\xi} T^{\mu\nu}(x)
\label{cal}
\end{equation}

The situation is completely different for the case of Eqs. (\ref{gh'}), 
(\ref{h'h}) and (\ref{LKK'}). Those transformations act only on the fields $\tilde{h}^{\mu\nu}$ and $K^{\alpha}_{\mu\nu}$, letting invariant the background metric $\gamma^{\mu\nu}(x)$. In this sense, one can interpret them
as true gauge transformations.
It can be shown that the dynamical equations for the gravitational field (\ref{G}) transform as a combination of themselves under the transformations (\ref{h'h}) and (\ref{LKK'}), supposing that the background space satisfies ${\stackrel{\circ}{R}}_{\mu\nu}  = 2\stackrel{\circ}{\Lambda}  \gamma_{\mu\nu}/(d-1)$. The new field $\tilde{h}'^{\mu\nu}(x)$ is also
a solution of the field equations, and
it corresponds to the same physical field as $\tilde{h}^{\mu\nu}(x)$.
In this case, the tensors do not transform in the usual way (\ref{cal}) but contains extra inhomogeneous terms which brings the possibility of annulling them. In this case, the energy-momentum tensor, for example, transforms according to: 
\begin{eqnarray}
T_{\mu\nu}(h',K') &=& T_{\mu\nu}(h,K)+
\frac{1}{16\pi G}\biggl\{ \hat{G}^{L}_{\mu\nu}\left[\frac{1}{\sqrt{-\gamma}}\sum^{\infty}_{k=1}
\frac{1}{k!}\pounds^{(k)}_{\xi}(\tilde{\gamma}^{\alpha\beta}+
\tilde{h}^{\alpha\beta})\right] \nonumber \\ & &
+ \gamma_{\alpha\mu}\gamma_{\beta\nu}
\frac{4\stackrel{\circ}{\Lambda}}{(d-1)\sqrt{-\gamma}}
\sum_{k=1}^{\infty}\frac{1}{k!}\pounds^{(k)}_{\xi}
(\tilde{\gamma}^{\alpha\beta}+\tilde{h}^{\alpha\beta})\biggr\},
\label{gauT}
\end{eqnarray}
where $\hat{G}^{L}_{\mu\nu}$ is the operator which when acting on $h^{\alpha\beta}$ yields expression (\ref{2G}). Hence, it is always possible to find gauge transformations (\ref{h'h}) and (\ref{LKK'}) which makes the energy-momentum tensor defined previously to be null. This is the analogue in the field theoretical approach to what happens with the pseudotensors in GR.
Note that the new energy-momentum tensor in Eq. (\ref{gauT}) 
is also covariantly conserved due to the properties of $G^{L}_{\mu\nu}$.

We have now completed our generalization of the field theoretical
approach for the case when the background metric satisfies Eq.
(\ref{1}), which includes the Ricci flat case when we make
$\stackrel{\circ}{\Lambda}  =0$. Although the energy-momentum tensor
defined above suffers from gauge ambiguities, the total energy or
angular momentum of a gravitational field does not. As we will see in
the next section, total conserved quantities are given by integration
of a $d$-form on a $d$-dimensional hypersurface, which can be reduced to
an integration of a $(d-1)$-form on the $(d-1)$-dimensional boundary of
this hypersurface at the asymptotic limit. As one would like to
preserve the asymptotic structure of the gravitational field, the gauge
vectors $\xi ^{\alpha}$ must satisfy boundary conditions at the
asymptotic $(d-1)$-dimensional surfaces. They must tend to AdS Killing
vectors at spatial infinity (see Ref. \cite{tei} for details), and the
true gauge transformations reduce to the identity at the boundaries.
Hence, as the total conserved quantities can be calculated by
integration on the asymptotic boundary, they are gauge independent.

\section{The $(d-1)$-form which gives total conserved quantities of asymptotically anti-de Sitter spacetimes}

Consider the energy-momentum tensor $T_{\mu\nu}$ defined in 
Eqs. (\ref{t}) and (\ref{tr}), and a Killing form $\xi _{\nu}$
of the background metric. Construct, as usual, the current
\begin{equation}
J^{\mu} = T^{\mu\nu}\xi _{\nu},
\end{equation}
and define the $d$-form
\begin{equation}
J = \frac{1}{d!}J^{\mu}\eta _{\mu \alpha_1...\alpha_d} dx^{\alpha _1}\wedge ...\wedge dx^{\alpha_d}, 
\label{dform}
\end{equation}
where ${\eta}_{\mu \alpha_1...\alpha_d} =
\sqrt{-\gamma} {\epsilon}_{\mu \alpha_1...\alpha_d}$ and 
${\epsilon}_{\mu \alpha_1...\alpha_d}$ is the $(d+1)$-dimensional (metric independent)
completely antisymmetric object.

Due to energy-momentum conservation together with the Killing equation, it
follows that $J^{\mu}_{\;\; ;\mu}=0$, which is equivalent to $dJ=0$, i.e., $J$ 
is a 
closed $d$-form. Hence,
\begin{equation}
\int _M dJ = \int _{\partial M} J = 0,
\label{int}
\end{equation}
where $M$ is a $(d+1)$-dimensional spacetime volume. 
If the boundary ${\partial M}$
of $M$ is constituted of two $d$-dimensional spacelike hypersurfaces
$\Sigma _1$ and $\Sigma _2$ labelled by the time parameters $t_1$ and
$t_2$, respectively, and a $d$ dimensional timelike hypersurface $B$ at
spatial infinity, and supposing that the fields and Killing vectors 
are such that $J$ is
zero at $B$ (which is the case for the non-radiating and asymptotically
AdS gravitational fields we will study in this paper), then Eq.
(\ref{int}) reduces to
\begin{equation}
\int _{\Sigma _2} J - \int _{\Sigma _1} J = 0,
\label{int2}
\end{equation}
and $\int _{\Sigma} J$ is a conserved quantity. If the Killing vector
field is timelike or associated with some rotational symmetry, then we
will have a conserved total energy or a conserved total angular
momentum, respectively.  Looking at the equations of motion (\ref{tT})
and the definition of $G^{L}_{\mu\nu}$ in Eq. (\ref{2G}), we can write
the conserved current $J^{\mu}$ on shell as:
\begin{equation}
J^{\mu} = \frac{1}{16\pi G}\biggr[
(\gamma^{\mu\nu}h^{\alpha\beta}+\gamma^{\alpha\beta}h^{\mu\nu}-
\gamma^{\alpha\mu}h^{\beta\nu}-\gamma^{\alpha\nu}h^{\beta\mu})_{;\alpha ;\beta}
\xi _{\nu} + \frac{4\stackrel{\circ}{\Lambda} }{d-1} h^{\mu\nu}\xi _{\nu}
\biggl].
\label{JT}
\end{equation}
Using the fact that the background metric satisfies
${\stackrel{\circ}{R}}_{\mu\nu}  = 2\gamma_{\mu\nu}
\stackrel{\circ}{\Lambda} /(d-1)$, the last term in the right-hand-side
of the above equation can be written in a suitable form:
\begin{equation}
\frac{4\stackrel{\circ}{\Lambda} }{d-1} h^{\mu\nu}\xi _{\nu}=
2(\xi_{\nu}h^{\nu}_{\lambda})^{[;\mu ;\lambda]},
\end{equation}
where $A^{[;\mu ;\lambda]}\equiv A^{;\mu ;\lambda} - A^{;\lambda ;\mu}$.
In this form the current $J^{\mu}$ does not depend anymore on $\stackrel{\circ}{\Lambda} $
and $d$. 

The following statement summarizes the mathematical achievement
of this work: the skew-symmetric tensor 
\begin{equation}
\Omega^{\alpha\mu} = \frac{1}{64\pi G}
(Q^{\alpha\mu\beta\nu}_{\;\;\;\;\;\;\; [;\beta}\xi_{\nu]}-
Q^{\alpha\mu\beta\nu}\xi_{\nu_;\beta}),
\label{om1}
\end{equation}
where
\begin{equation}
Q^{\alpha\mu\beta\nu} \equiv
\gamma^{\mu\nu}h^{\alpha\beta}+\gamma^{\alpha\beta}h^{\mu\nu}-
\gamma^{\alpha\nu}h^{\beta\mu}-\gamma^{\beta\mu}h^{\alpha\nu},
\label{om2}
\end{equation}
satisfies
\begin{equation}
J^{\mu}=2\Omega^{\mu\alpha}_{\;\;\;\; ;\alpha} .
\label{dom}
\end{equation}
The tensor $Q^{\alpha\mu\beta\nu}$ has the same symmetries as the Riemman tensor.
Defining the $(d-1)$-form 
\begin{equation}
\Omega = \frac{1}{(d-1)!}\Omega^{\mu\nu}\eta _{\mu\nu \alpha_1...\alpha_{d-1}} dx^{\alpha _1}\wedge ...\wedge dx^{\alpha _{d-1}},
\label{om3}
\end{equation}
then Eq. (\ref{dom}) is equivalent to
\begin{equation}
J=d\Omega,
\end{equation}
which proves that $J$ is not only closed, but globally exact. Hence, the
conserved quantities will be given by
\begin{equation}
Q = \int_{\partial \Sigma} \Omega .
\label{om4}
\end{equation}

This total conserved quantity is gauge independent
because it was reduced to an integral at the boundary at infinity.
There, the true gauge tranformations (\ref{h'h}) reduce to the identity
because we are imposing the preservation of the asymptotic structure of
the fields. Also, with the $(d-1)$-form $\Omega$ we do not need the knowledge
of the gravitational field on the whole $\Sigma$ but only its asymptotic
behaviour on $\partial\Sigma$, which makes the calculation of the 
total conserved quantities much easier and general\footnote{In Ref.
\cite{gri2} another energy-momentum tensor is defined for flat
background, and Eq. (\ref{tT}) is written in another form, with the
presence of extra non linear terms in its left-hand-side. However, as
total conserved quantities are calculated by means of asymptotic
boundary integrals, only the linear terms given by $\Omega$ in Eq.
(\ref{om4}) are important.}.

Note that $\Omega$ does not depend neither on $d$ nor on $\Lambda$.
All dependence on these parameters are contained in $h^{\mu\nu}$. However,
we have to fix the background metric in some way because once a metric
$g_{\mu\nu}$ is given, the determination of $h^{\mu\nu}$ will depend on
the choice of $\gamma_{\mu\nu}$, as can be seen from Eq. (\ref{gh}).
Note that even for asymptotically AdS spacetimes one could take
Ricci flat background geometries just by making $\stackrel{\circ}{\Lambda}=0$ 
in Eq. (\ref{1}): our results are independent on the choice of
$\stackrel{\circ}{\Lambda}$.
Let us then calculate the total energy of a simple asymptotically
AdS spacetime, the Schwarzschild AdS solution in four dimensions, 
using a Ricci flat background,
namely, a flat background. The Schwarzschild AdS solution is:
\begin{equation}
\label{21}
ds^{2} = 
-\biggl[1 + \biggl(\frac{r}{R}\biggr)^{2} -\frac{2m}{r}\biggr] dt^{2} + 
\biggl[1 + \biggl(\frac{r}{R}\biggr)^{2} - \frac{2m}{r}\biggr]^{-1}
dr^{2} + r^2[d\theta^2 + \sin^2(\theta)d\phi^2]
\end{equation}
where $R$ is the radius of curvature of such space, related to the
effective cosmological constant $\Lambda _{eff} =\Lambda + 8\pi G \rho_V$ by 
$R = (-3/\Lambda _{eff})^{1/2}$ ($\rho _V$ is the matter vacuum energy density).
The flat background metric is taken in spherical coordinates.
The non null $h^{\mu\nu}$ are
\begin{equation}
\label{22}
h^{tt}=\frac{r\rho^2-2m}{r-2m+r\rho^2},
\end{equation}
and
\begin{equation}
\label{23}
h^{rr}=\frac{-2m}{r}+\rho^2,
\end{equation}
where $\rho\equiv r/R$.
To calculate the total energy, we will take the timelike Killing
vector field of flat spacetime, $\xi^{\mu}=\delta^{\mu}_{t}$.
When we calculate the total energy using Eqs. (\ref{om1}), (\ref{om2}),
(\ref{om3}) and (\ref{om4}), we obtain (from now on we will make $G=1$):
\begin{equation}
\label{24}
E=\lim_{r \rightarrow \infty} \biggr[\frac{r^2(m+r\rho^2)}{2(r-2m+r\rho^2)^2}
+\frac{m}{2}-\frac{r^3}{R^2}\biggl] = \frac{m}{2}-
\lim_{r \rightarrow \infty}\frac{r^3}{R^2}.
\end{equation}
The last term is the energy of the anti-de Sitter solution, which
is the Schwarzschild AdS solution with $m=0$. Subtracting it we find:
\begin{equation}
\label{25}
E_{ren}=\frac{m}{2}.
\end{equation}
This result does not give the Schwarzschild mass, first calculated in
the field theoretical approach in Ref. \cite{gri}, when we put 
$\Lambda _{eff}$ to zero after the calculation. For the usual asymptotically
flat Schwarzschild solution, the energy in Eq.  (\ref{24}), setting
$\Lambda _{eff}=\rho=0$ from the beginning, reads $E=m$. Hence, we have
an internal inconsistency: the total energy of the Schwarzschild AdS
solution using a flat background does not yield the total energy of the
asymptotically flat Schwarzschild solution (calculated within the same
rules) when we put $\Lambda _{eff}$ to zero. However, as we will see in
the following, if we take an ADS background (with the same 
$\Lambda _{eff}$) in the calculation of the energy of the Schwarzschild AdS
solution, the energy in the limit $\Lambda _{eff} =0$ yields the energy
of the asymptotically flat Schwarzschild solution calculated in Ref.
\cite{gri}. This shows that the background metric is not arbitrary but
is dictated by the asymptotic structure of the spacetime in question.
If we take other simple examples, we can readily conclude that the
background metric for asymptotic AdS spacetimes must not only satisfy
${\stackrel{\circ}{R}}_{\mu\nu}  = 2\stackrel{\circ}{\Lambda}
\gamma_{\mu\nu} /(d-1)$, with $\stackrel{\circ}{\Lambda} = \Lambda
_{eff} = \Lambda + 8\pi G \rho_V$, but it must also be asymptotically
AdS. A metric which is asymptotically AdS, satisfies Eq. (\ref{1}), and
is regular everywhere must be the AdS metric (an analogous reasoning
can be used for asymptotically flat spacetimes). Hence, the asymptotic
structure, together with regularity assumptions, fixes the background
metric.  Also, the background AdS metric must be in the same coordinate
system as the AdS asymptotic geometry at infinity, but this coordinate
system may be arbitrary because $\Omega^{\alpha\mu}$ in Eq. (\ref{om1})
is a true tensor. This is equivalent to the imposition that only true
gauge transformations which reduce to the identity at the spatial
infinity are allowed. With these restrictions dictated by the
asymptotic structure of the geometry under study, the conserved
quantity (\ref{om4}) has no ambiguities: it is invariant under the
allowed true gauge transformations, and the background metric is fixed
by the asymptotic behaviour of the geometry.

Note that the condition $\stackrel{\circ}{\Lambda} = \Lambda _{eff}
= \Lambda + 8\pi G \rho_V$ is also the constraint one must impose on 
$\stackrel{\circ}{\Lambda}$ in order to have $h_{\mu\nu}=0$ as a
solution of the vacuum field equations (\ref{K}). 

With these rules in mind, let us calculate some conserved quantities
for asymptotically AdS solutions. In the examples below we will take for 
convenience, and without loss of generality, that $\rho_V=0$ and hence
$\stackrel{\circ}{\Lambda} =\Lambda$. Also, all the quantities calculated
below are conserved because in all cases $\int _{B} J = 0$, where
$B$ is the timelike portion of $\partial M$ in Eq. (\ref{int}),
from where Eq. (\ref{int2}) follows.

\begin{enumerate}
\item {\bf The Kerr-anti-de Sitter spacetime.}

The coordinates of the effective geometry $g_{\mu\nu}$ are such that it 
tends asymptotically to the anti-de Sitter metric in the form
\begin{equation}
\label{ads1}
d{\stackrel{\circ}s}^{2} = -\biggl[1 + \biggl(\frac{r}{R}\biggr)^{2}\biggr] dt^{2} + \biggl[1 + \biggl(\frac{r}{R}\biggr)^{2}\biggr]^{-1}
dr^{2} + r^2 [d\theta^2 + \sin^2(\theta)d\phi^2]
\end{equation}
where, as before, $R = (-3/\Lambda)^{1/2}$. 
In this case, the leading order aymptotic components of the gravitational 
field $h^{\mu\nu}$ read \cite{tei}
\begin{equation}
\label{80}
\left\{
\begin{array}{rcl}
h^{tt} &=&-\frac{2mR^4}{r^5} [1 - {\alpha}^2 {\sin}^{2}{\theta}]^{-5/2} \\
h^{t\phi} &=& -\frac{2amR^2}{r^5} [1 - {\alpha}^2 {\sin}^{2}{\theta}]^{-5/2} \\
h^{\phi\phi} &=&-\frac{2ma^2}{r^5} [1 - {\alpha}^2 {\sin}^{2}{\theta}]^{-5/2} \\
h^{rr} &=& - \frac{2m}{r} [1 - {\alpha}^2 {\sin}^{2}{\theta}]^{-3/2} \\
h^{r\theta} &=& \frac{2ma^2}{r^4} [1 - 
{\alpha}^2 {\sin}^{2}{\theta}]^{-5/2} \sin{\theta} \cos{\theta} \\
h^{\theta\theta} &=& - \frac{2ma^4}{r^7} [1 - {\alpha}^2 {\sin}^{2}{\theta}]^{-7/2}
{\sin}^{2}{\theta} {\cos}^{2}{\theta} \\
\end{array}
\right.
\end{equation}
where $\alpha = a/R$, and $a$ is related to the angular momentum
per unit mass of the Kerr-anti-de Sitter spacetime.

For the calculation of the gravitational energy, we take the timelike
Killing vector field of the AdS spacetime
$\xi^{\mu}=\delta^{\mu}_{t}$, which is also a Killing vector field of
the effective geometry, and insert it in Eq.  (\ref{om1}). The only nonzero
component of $\Omega^{\alpha\mu}$ is $\Omega^{tr}$, which yields for
$Q$ in Eq. (\ref{om4})
\begin{equation}
\label{17}
Q = E = \frac{m}{(1 - {\alpha}^2)^2},
\end{equation}
agreeing with the results using pseudotensors \cite{des}, the
hamiltonian formalism \cite{tei}, and the quasilocal stress tensor
\cite{ids}. Note that the flat space limit $\Lambda = 0$ in the above
expression yields the energy of the asymptotically flat Kerr spacetime
calculated within the same rules. There are no internal
inconsistencies.

The Kerr-anti-de Sitter spacetime also has a conserved angular momentum.
To calculate it, we now have to take the Killing vector field
$\xi^{\mu}=\delta^{\mu}_{\phi}$, which is again a Killing vector field of
the effective geometry. We proceed in an analogous way obtaining
\begin{equation}
\label{18}
Q = L_{\phi} = -\frac{ma}{(1 - {\alpha}^2)^2}
\end{equation}
which coincide with calculations on Refs. \cite{tei,ids}

\item {\bf The Schwarzschild-anti-de Sitter spacetime in five dimensions.}

The Schwarzschild-anti-de Sitter metric in five dimensions reads
\begin{eqnarray}
\label{s5}
ds^{2}&=&-\biggl[1 + \biggl(\frac{r}{R}\biggr)^{2} - \biggl(\frac{r_0}{r}\biggr)^{2}\biggr] dt^{2} + \biggl[1 + \biggl(\frac{r}{R}\biggr)^{2} - \biggl(\frac{r_0}{r}\biggr)^{2}\biggr]^{-1}dr^{2}
\nonumber \\& & +
r^2 \{d\theta_1^2 + \sin^2(\theta_1)[d\theta_2^2 + \sin^2(\theta_2)d\phi^2]\},
\end{eqnarray}
where $R = (-6/\Lambda)^{1/2}$, and the background metric is obtained from
Eq. (\ref{s5}) by making $r_0=0$ in it. The non null $h^{\mu\nu}$ are
\begin{equation}
\label{225}
h^{tt}=-\frac{l^2}{(1+\rho^2)(1+\rho^2-l^2)},
\end{equation}
and
\begin{equation}
\label{235}
h^{rr}=l^2,
\end{equation}
where $l\equiv r_0/r$ and, as before, $\rho\equiv r/R$.
Taking again the timelike
Killing vector field of the $AdS$ spacetime
$\xi^{\mu}=\delta^{\mu}_{t}$, and inserting it in Eq. (\ref{om1}), 
we obtain for the total
the gravitational energy of this spacetime the value
\begin{equation}
\label{175}
Q = E = \frac{3\pi r_0^2}{8}
\end{equation}
which agree with the result of Ref. \cite{mye2}. This is the only
non null conserved quantity of this geometry.

\item {\bf The near-horizon limit of the D3-brane}

The effective five-dimensional metric now reads
\begin{equation}
\label{D3}
ds^{2} = \biggl(\frac{r}{R}\biggr)^{2} \biggl\{- \biggl[1-\biggl(\frac{r_0}{r}\biggr)^{4}\biggr] dt^{2} + (dx^i)^2\biggr\} + 
\biggl[1 - \biggl(\frac{r_0}{r}\biggr)^{4}\biggr]^{-1}\biggl(\frac{r}{R}\biggr)^{2}dr^{2}.
\end{equation}
The background metric is obtained by making $r_0=0$ in 
Eq. (\ref{D3}). It is now written in a different
coordinate system than in the precedent example but in accordance
with the asymototic limit of (\ref{D3}).
The non null $h^{\mu\nu}$ are
\begin{equation}
\label{225D}
h^{tt}=-\frac{l^4}{\rho^2(1-l^4)},
\end{equation}
and
\begin{equation}
\label{235D}
h^{rr}=l^4\rho^2,
\end{equation}
where $l$ and $\rho$ are defined as above.
Taking again the timelike
Killing vector field of the AdS spacetime,
the gravitational energy now reads
\begin{equation}
\label{175D}
Q = E = \frac{3r_0^4}{16\pi R^5}\int d^3x,
\end{equation}
which agree with the result of Ref. \cite{mye2}. This is the only
non null conserved quantity of this geometry.

\end{enumerate}

\section{Conclusion}

In this paper we have extended the field theoretical approach to GR to
the case where the background metric satisfies
${\stackrel{\circ}{R}}_{\mu\nu}  =
2\gamma_{\mu\nu}{\stackrel{\circ}{\Lambda}} /(d-1)$. After that, we
have obtained a $(d-1)$-form $\Omega$ which, when integrated on
asymptotic $(d-1)$-dimensional surfaces, yields the values of total
energies and angular momenta of asymptotically $(d+1)$-dimensional AdS
or flat spacetimes.  Although the dynamics of the effective geometry
does not depend on the background metric we choose, the values of those
total conserved quantities are strongly affected by the choice we
make.  As we have shown in the text, if we do not choose judiciously
the background metric we may obtain preposterous results for the
gravitational energy.  Hence, the ambiguity in the choice of the
background metric may be eliminated only by going beyond the equations
of motion and examining further concepts, like the consistency of
calculations of total energy and angular momenta. These considerations
indicate,
together with regularity conditions, what is the background metric one
should adopt.

The calculations of total conserved quantities using $\Omega$ yield
finite results, and are gauge independent once one does not violate
asymptotic conditions. This is not true, however, for conserved
quantities contained in finite regions of the background space. These
calculations may suffer from gauge ambiguities because the effective
geometry may be presented in many different coordinate systems with the
same asymptotic limits, and hence, in finite regions, the true gauge
transformations (\ref{h'h}) are not trivial. It should be interesting
to investigate under what subgroup of the true gauge transformations
(\ref{h'h}) is the $(d-1)$-form $\Omega$ given in Eqs. (\ref{om1}) and
(\ref{om3}) invariant.

Let us now compare the $(d-1)$-form $\Omega$ given in Eqs. (\ref{om1},\ref{om3}) and its integral with the quasilocal stress tensor of
Ref. \cite{bro}, and the surface integrals of Ref. \cite{tei}. They
have in common the presence of a background (reference) space which is
fixed by the asymptotic behaviour of the effective geometry. However,
for the $(d-1)$-form $\Omega$, the presence of the background space is
much more important. Contrary to the other prescriptions, the surfaces
where the integrals are performed are defined on the background space,
and the Killing vector fields which are present in $\Omega$ generates
isometries of the background, not of the effective geometry.  This last
fact does not mean that we can have more conserved quantities than the
number of Killing vectors of the effective geometry. If we use some
Killing vector field of the background geometry which does not describe
an isometry of the effective geometry than the integral $\int _{B} J$,
where $B$ is the timelike portion of $\partial M$ in Eq. (\ref{int}),
is not zero and Eq. (\ref{int2}) does not follow. The quantity 
$\int _{\Sigma} J$ is not conserved because there is a flux of $J$ 
through $B$.

The $(d-1)$-form $\Omega$ can be used to calculate conserved quantities
or fluxes through the boundary $B$ for more involved effective
geometries.  The fact that $\Omega$ is derived from a theory which
describes the gravitational field as a spin-two field propagating on a
fixed background may be useful to understand some aspects of the
correspondence of conformal field theory in an AdS boundary and
gravitational theory in AdS spaces.

{\bf ACKNOWLEDGEMENTS}

We would like to thank the Cosmology Group of CBPF for useful discussions, 
and CNPq of Brazil for financial support.


\begin{thebibliography}{99}

\bibitem{mal} J. M. Maldacena, Adv. Theor. Math. Phys. {\bf 2}, 231 (1997).
\bibitem{wit} E. Witten, Adv. Theor. Math. Phys. {\bf 2}, 253 (1998).
\bibitem{des} L. F. Abbott and S. Deser, Nucl. Phys. {\bf B 195}, 76
(1982).
\bibitem{tei} M. Henneaux and C. Teitelboim, Commun. Math.
Phys. {\bf 98}, 391 (1985).
\bibitem{ids} N. Pinto-Neto and I. Dami\~ao Soares, Phys. Rev. {\bf D 52}, 
5665 (1995).
\bibitem{kra} V. Balasubramanian and P. Kraus, hep-th/9902121.
\bibitem{mye} R. C. Myers, hep-th/9903203.
\bibitem{des2} S.Deser, GRG, {\bf 1}, 9 (1970).
\bibitem{gri} L.P.Grishchuk, A.N. Petrov and A.D. Popova, Commun. 
Math. Phys. 94, 379 (1984).
\bibitem{gri2} S. V. Babak and L.P.Grishchuk, gr-qc/9907027.
\bibitem{ron} R. R. Silva, J. Math. Phys., {\bf 39}, 6206 (1998).
\bibitem{mye2} G. T. Horowitz and R. C. Myers,  Phys. Rev. {\bf D 59},
026005 (1999).
\bibitem{bro} J. D. Brown and J. W. York, Jr., Phys. Rev. {\bf D 47},
1407 (1993).



\end{thebibliography}
\end{document}